\documentclass[aps,prl,twocolumn,showpacs,superscriptaddress,groupedaddress]{revtex4}  
\usepackage{graphicx}  
\usepackage{dcolumn}   
\usepackage{bm}        
\usepackage{amssymb}   
\usepackage{amsmath}
\usepackage{float}
\usepackage{xcolor}

\begin{document}



\title{Scalings in Coalescence of Liquid Droplets}

\author{Xi Xia}
\author{Chengming He}
\author{Peng Zhang}%
\email{pengzhang.zhang@polyu.edu.hk}
\affiliation{
Department of Mechanical Engineering, The Hong Kong Polytechnic University, Hung Hom, Hong Kong}
\date{\today}

\begin{abstract}
This letter presents a scaling theory of the coalescence of two viscous spherical droplets. An initial value problem was formulated and analytically solved for the evolution of the radius of a liquid neck formed upon droplet coalescence. Two asymptotic solutions of the initial value problem reproduce the well-known scaling relations in the viscous and inertial regimes. The viscous-to-inertial crossover experimentally observed by Paulsen \textit{et al.} [Phys. Rev. Lett. {\bf 106}, 114501 (2011)] manifests in the theory, and their fitting relation, which shows collapse of data of different viscosities onto a single curve, is an approximation to the general solution of the initial value problem.
\end{abstract}

\maketitle


Droplet coalescence is a ubiquitous phenomenon in natural and industrial processes that involve dispersed two-phase flows \cite{Yarin2006,Gorokhovski2008,Kavehpour2015,Low1982,Chen2007,Derby2010,Tang2016,Xia2017,vandeVorst1994,Dreher1999,Squires2005}. Among the various aspects of droplet coalescence, the initial coalescence of two liquid droplets has been of core interest. The first quantitative analysis of sphere coalescence was provided by Frenkel \cite{Frenkel1945} based on the assumption of internal Stokes flow; however, the result was commented as ``misleading" by Hopper \cite{Hopper1993a}, who gave an analytical two-dimensional solution for the coalescence of two cylindrical droplets \cite{Hopper1993a,Hopper1990,Hopper1993b} for viscous sintering. He showed that the non-dimensional radius $R^{\ast}$ (scaled by $R_0$, the initial radius of the droplets) of the neck (or bridge) between the droplets grows as $R^{\ast} \sim t^{\ast} \ln{t^{\ast}}$ with $t^{\ast}$ being the time scaled by $\tau_v=\mu R_0/\sigma$, where $\mu$ is the dynamic viscosity and $\sigma$ the surface tension coefficient. This scaling law was later extended by Eggers \textit{et al.} \cite{Eggers1999} to the three-dimensional coalescence in the very early stage when $R^{\ast} \ll 1$. For larger $R^{\ast}$, they \cite{Eggers1999,Duchemin2003} argued that the neck flow goes beyond the Stokes regime to the inertial (or inviscid) regime, and further arrived at the $1/2$ power-law scaling, $R^{\ast} \sim (t^{\ast})^{1/2}$ with the time scale being $\tau_i=(\rho R_0^3/\sigma)^{1/2}$, where $\rho$ is the liquid density.

Recent advances in the fast digital imaging \cite{Aarts2005,Thoroddsen2005,Yao2005}, state-of-art probing techniques \cite{Case2008,Fezzaa2008,Paulsen2011}, and numerical simulation enabled researchers to scrutinize the early stages of drop coalescence, roughly corresponding to $0<R^{\ast}<1$. As a result, the $1/2$ power-law scaling was confirmed by many experimental \cite{Aarts2005,Thoroddsen2005,Fezzaa2008,Wu2004,Burton2007,Case2009} and numerical \cite{Duchemin2003,Eiswirth2012,Pothier2012,Gross2013,Sprittles2012} studies, with the inertial time scale identified to be exactly $\tau_i$. The same scaling was also verified for droplet coalescence on substrate \cite{Ristenpart2006,Sanchez2012,Eddi2013}. However, the studies of Aarts \textit{et al.} \cite{Aarts2005} and Thoroddsen \textit{et al.} \cite{Thoroddsen2005} pointed to a glaring fact that the viscous regime is better predicted by the linear scaling of $R^{\ast} \sim t^{\ast}$ rather than the scaling of $R^{\ast} \sim t^{\ast}\ln{t^{\ast}}$. This linear correlation was also corroborated by several studies \cite{Yao2005,Paulsen2011,Burton2007}.

Subsequently, researchers started to direct their attention towards the crossover (or transition) between the viscous and inertial regimes. The first direct evidence of the crossover from $R^{\ast} \sim t^{\ast}$ to $R^{\ast} \sim (t^{\ast})^{1/2}$ was reported by Burton and Taborek \cite{Burton2007}. By equating the characteristic velocities from the two scaling laws, they derived the crossover length, $l_c \sim \mu(R_0/\rho \sigma)^{1/2}$, which is consistent with the results of Paulsen and coworkers \cite{Paulsen2011,Paulsen2013}, who obtained the same crossover length and, additionally, the crossover time, $\tau_c \sim \mu^2(R_0/\rho \sigma^3)^{1/2}$, by assuming unity Reynolds number. Applying these time and length scales, Paulsen and coworkers collapsed the neck evolutions of distinct viscosities onto a single fitting curve, $(R^{\ast})^{-1} \sim (t^{\ast})^{-1}+(t^{\ast})^{-1/2}$, indicating a certain degree of universality for droplet coalescence. 

\begin{figure}
\includegraphics[scale=0.4]{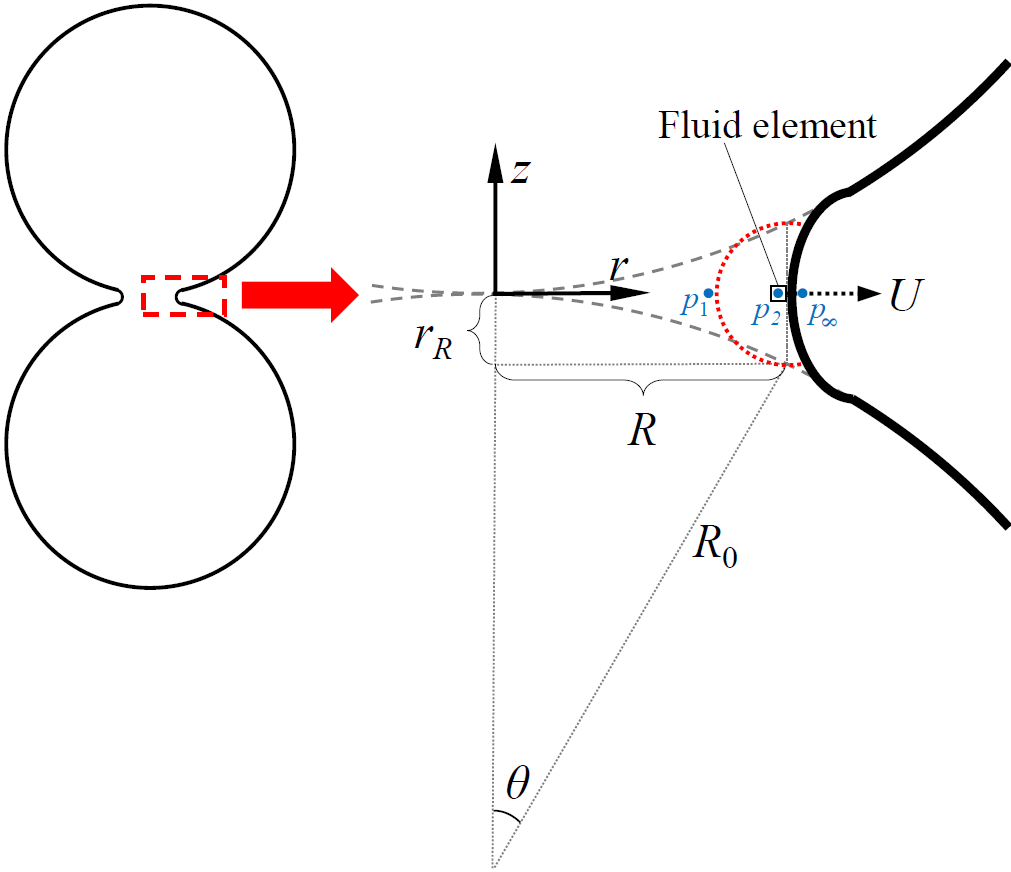}
\caption{A zoomed-in schematic of the neck region between two merging droplets. The red dotted curve together with the droplet interface encircles the main flow region of the droplet, outside of which the flow may be considered to be negligible.}
\label{fig:1}
\end{figure}

In this Letter, we take one step further to theoretically analyze the neck evolution using the Navier-Stokes equation. The very beginning of droplet contact, where the length scale of the neck is comparable to the mean free path of the gas medium, is beyond the scope of the current theory. 
A schematic of the neck between two merging droplets of initial radius $R_0$ is shown in Fig.~\ref{fig:1}. The neck radius, $R$, is defined as the minimum distance of the neck from the $z$-axis. It follows from the geometry that the characteristic width (or height) of the neck is approximately $2r_R$, with $r_R = R \tan{(\theta/2)}$. The open square dot represents an infinitesimal fluid element locating on the r-axis and next to the droplet neck surface; thus, it moves with the center of the neck at the same speed of $U$. 
%

Given axisymmetric flow without swirling, the dynamics of the fluid element is governed by the $r$-direction N-S equation,
\begin{equation}
\begin{split}
\label{eq:2}
& \rho (\partial_t u_r + u_z \partial_z u_r + u_r \partial_r u_r)\\
& = -\partial_r p + \mu \left(\partial_z^2 u_r + \partial_r^2 u_r + \frac{\partial_r u_r}{r} - \frac{u_r}{r^2} \right),
\end{split}
\end{equation}
where $u_z$ and $u_r$ are the velocity components in the $z$- and $r$- directions, respectively; $p$ is the pressure and $t$ the time. We now list the physical assumptions and approximations that are necessary for further derivations.\\ 
(i) The flow is localized around the neck region. This means that the neck movement only affects its immediate near field encircled by the red dotted curve shown in Fig.~\ref{fig:1}, whereas the rest of the drop is considered nearly stagnant. This assumption is in line with the finding of Paulsen \textit{et al.} \cite{Paulsen2011} that the flow extends over a length comparable to the neck height rather than the neck radius. We also note that this assumption implies $R \ll R_0$ so that the rest of the droplets is hardly ``felt" by the neck.\\
(ii) The flow is quasi-steady \cite{Eggers1999,Duchemin2003,Gross2013}, meaning $\partial_t u_r \approx 0$.\\
(iii) The neck geometry is associated with two length scales, $R$ and $r_R$, corresponding to two principle curvatures, $1/R$ and $1/r_R$. Therefore, the neck curvature in the $zr$-plane is approximately $1/r_R$ \cite{Gross2013}.

Since the fluid element in Fig.~\ref{fig:1} moves along the $r$-axis, $u_z$, $\partial_z u_r$, and $\partial_z u_z$ should all be zero at $z=0$ owing to the condition of symmetry, so the term $u_z \partial_z u_r$ vanishes in Eq.~(\ref{eq:2}). Applying the continuity equation and the relation $\partial_z u_z = 0$, we have $\partial_r u_r = -u_r/r$ and then $\partial_r^2 u_r = 2u_r/r^2$. According to Assumption (i), the length scale of the encircled flow region is $r_R$ in both $z$- and $r$- directions, so $\partial_r u_r = -\partial_z u_r \sim U/r_R$. The following scaling approximations can be readily obtained: $u_r \partial_r u_r \sim U^2/r_R$, $\partial_r p \sim \Delta p/r_R$, $\partial_z^2 u_r \sim -U/r_R^2$, $\partial_r u_r/r \sim U/(R r_R)$, where $\Delta p$ is the pressure difference across the encircled flow region. Consequently, with $(u_r/r^2)|_{r=R} = U/R^2$, Eq.~(\ref{eq:2}) takes the simplified form,
\begin{equation}
\label{eq:3}
\frac{\rho U^2}{r_R} = -\frac{C_1 \Delta p}{r_R} - \frac{C_2 \mu U}{r_R^2} + \frac{C_3 \mu U}{Rr_R} + \frac{\mu U}{R^2},
\end{equation}
where $C_1$, $C_2$, and $C_3$ are positive scaling coefficients to be determined experimentally. Note that the derivation of Eq.~(\ref{eq:3}) also requires Assumption (ii). $\Delta p$ can be estimated from $\Delta p = p_2-p_1$, where $p_1$ and $p_2$ are the pressures corresponding to the locations shown in Fig.~\ref{fig:1}. As the present study concerns the coalescence of liquid droplets in a gaseous environment, the liquid-gas interface can be considered as a free surface, where the capillary pressure jump is given by $p_\infty - p = -2\mu \boldsymbol{n}\cdot \boldsymbol{S} \cdot \boldsymbol{n} + \sigma\kappa$ \cite{Tryggvason2011}, where $p_\infty$ is the ambient gas pressure,  $\boldsymbol{n}$ and $\kappa$ are the unit normal vector and curvature of the interface, respectively, and $\boldsymbol{S}$ is the rate-of-strain tensor. It is noted that the static version of this condition is known as the Young-Laplace equation. Accordingly, the pressure jumps at the interfaces of the far-side droplet and the neck satisfy $p_\infty - p_1 = -2\sigma/R_0$ and $p_\infty - p_2 = -2\mu\partial_r u_r+\sigma(1/r_R - 1/R)$, respectively, which can be subtracted to yield $\Delta p = -\sigma(1/r_R - 1/R + 2/R_0) - 2\mu U/R$, where $(\partial_r u_r)|_{r=R}=-U/R$ from the continuity equation is applied. With $R \ll R_0$ implied by Assumption (i), $\theta \approx \sin{\theta} = R/R_0$ and $\theta/2 \approx \tan{(\theta/2)} = r_R/R$, it gives the geometric relationship,
\begin{equation}
\label{eq:4}
\frac{r_R}{R} \approx \frac{R}{2R_0},
\end{equation}
which is consistent with previous studies \cite{Paulsen2011,Gross2013}. Based on Eq.~(\ref{eq:4}), we can make the approximation that $\Delta p \approx -\sigma/r_R$. With $r_R \ll R$, the third and fourth terms on the right-hand side of Eq.~(\ref{eq:3}) are negligible compared with the second term. Together with $\dot{R} = \mathrm{d}R/dt = U$, Eq.~(\ref{eq:3}) can be therefore simplified into
%
%
%
\begin{equation}
\label{eq:6}
\frac{\rho \dot{R}^{\ast 2} L^2}{T^2} = \frac{C_1 \sigma D_0}{R^{\ast 2}L^2} - \frac{C_2 \mu D_0 \dot{R}^{\ast}}{R^{\ast 2}LT},
\end{equation}
where $D_0=2R_0$, $R^{\ast} = R/L$, $\dot{R}^{\ast}=\dot{R}/U$, and $T=L/U$, with $L$, $U$, and $T$ being the characteristic length, velocity, and time scales, respectively. Eq.~(\ref{eq:6}) describes a balance among the effects of advection, pressure gradient, and diffusion. 

The experimental studies of Paulsen and coworkers \cite{Paulsen2011,Paulsen2013} imply the existence of a unified formulation given the length and time are scaled properly. If Eq.~(\ref{eq:6}) is such a formulation, we must have 
\begin{equation}
\label{eq:7}
\frac{\rho L^2}{T^2} = \frac{\sigma D_0}{L^2} = \frac{\mu D_0}{LT},
\end{equation}
yielding $L = Oh D_0$ and $T = \mu Oh D_0/\sigma$, where $Oh = \mu/\sqrt{\rho \sigma D_0}$ is the Ohnesorge number. 
Note that $L$ and $T$ match exactly with the viscous-to-inertial crossover scales used by Burton and Taborek \cite{Burton2007} and Paulsen and coworkers \cite{Paulsen2011,Paulsen2013}. 

Now, Eq.~(\ref{eq:6}) takes the dimensionless form,
\begin{equation}
\label{eq:8}
\dot{R}^{\ast 2} = \frac{C_1}{R^{\ast 2}} - \frac{C_2 \dot{R}^{\ast}}{R^{\ast 2}}.
\end{equation}
Bearing in mind that $C_1$ and $C_2$ are positive and of $O(1)$, we can integrate Eq.~(\ref{eq:8}) with the initial condition $R^{\ast}(t^{\ast}=0) = 0$, where $t^{\ast} = t/T$, to obtain the exact solution,
\begin{equation}
\label{eq:9}
\begin{split}
t^{\ast} & = \frac{C_2 R^{\ast}}{2C_1} + \frac{C_2}{4C_1}\left[R^{\ast}\sqrt{\frac{4C_1 R^{\ast 2}}{C_2^2} + 1} \right.\\
& \left.+ \frac{C_2}{2\sqrt{C_1}}\sinh^{-1}\left(\frac{2\sqrt{C_1} R^{\ast}}{C_2}\right) \right].
\end{split}
\end{equation}
Eq.~(\ref{eq:9}) readily dictates the asymptotic behaviors associated with the viscous and inertial regimes. For $R^{\ast} \gg C_2/\sqrt{4C_1} = O(1)$, Eq.~(\ref{eq:9}) yields
\begin{equation}
\label{eq:10}
t^{\ast} \approx \frac{R^{\ast 2}}{2\sqrt{C_1}} + O(R^{\ast}).
\end{equation}
Eq.~(\ref{eq:10}) can be rewritten in the dimensional form of $R/R_0 \approx c_1(t/\tau_i)^{1/2}$, with $c_1=(8C_1)^{1/4}$. It is in line with the $1/2$ power-law scaling for the inertial coalescence regime. For $R^{\ast} \ll C_2/\sqrt{4C_1}$, Eq.~(\ref{eq:9}) yields 
\begin{equation}
\label{eq:11}
\begin{split}
t^{\ast} & \approx \frac{C_2}{2C_1}\left[\frac{3R^{\ast}}{2} + \frac{C_2}{4\sqrt{C_1}}\ln\left(\frac{2\sqrt{C_1} R^{\ast}}{C_2} + 1\right) \right]\\
& \approx \frac{C_2 R^{\ast}}{C_1} + O(R^{\ast 2}),
\end{split}
\end{equation}
which can also be expressed in the dimensional form of $R/R_0 \approx c_2 t/\tau_v$, with $c_2=C_1/C_2$. 
It is noted that Eq.~(\ref{eq:11}) can be also reduced to the form of $R \sim t\sigma/\mu$, which is void of any characteristic length. This can be interpreted that the physics of viscous regime is intermediate self-similar \cite{Barenblatt1996}. The coefficients $c_1 = 1.68$ and $c_2 = 1$ were suggested by Paulsen \cite{Paulsen2013} although different values were reported \cite{Aarts2005,Thoroddsen2005,Wu2004}. Accordingly, we obtain $C_1 \approx 1$ and $C_2 \approx 1$, which confirms that these coefficients are of $O(1)$. 

Here, we contrast our viscous scaling with the $t^*\ln(t^*)$ behavior given by Eggers \textit{et al.} \cite{Eggers1999}. In fact, our supplementary materials \cite{SM} now confirm the validity of their scaling through a detailed derivation of a key velocity integral, which was only provided via an argument in Eggers \textit{et al.}'s original work. It is noted that while their scaling is based on the accurate solution of Stokes flow and should be valid for $t \longrightarrow 0$, it could divert from the actual neck evolution. For one thing, the initial merging could start from a non-zero gap height, so that their requirement for $r_R$ being two-order-of-magnitude smaller than $R$ would not be satisfied at the beginning. Furthermore, as the neck expands out, the shape of the meniscus evolves in both $z$ and $r$ directions, which causes the surface tension to be more distributed rather than concentrated. Therefore, their approximation of parallel undisturbed interfaces on both sides of the neck and the treatment of applying ring force or belt force could introduce additional error. Our approach, on the hand, can be understood as a linearization of the N-S equation in the vicinity of the neck so the $\ln(t^*)$ term is smoothed out to be a constant within a finite time period. The merit is that both viscous and inertial terms can be preserved to yield a unified formula for a significantly extended lifespan of the neck evolution, covering the viscous, viscous-to-inertial crossover, and inertial regimes. The performance of our theory will be justified through comparisons with experiment in the following.
%
\begin{figure}
\includegraphics[scale=0.37]{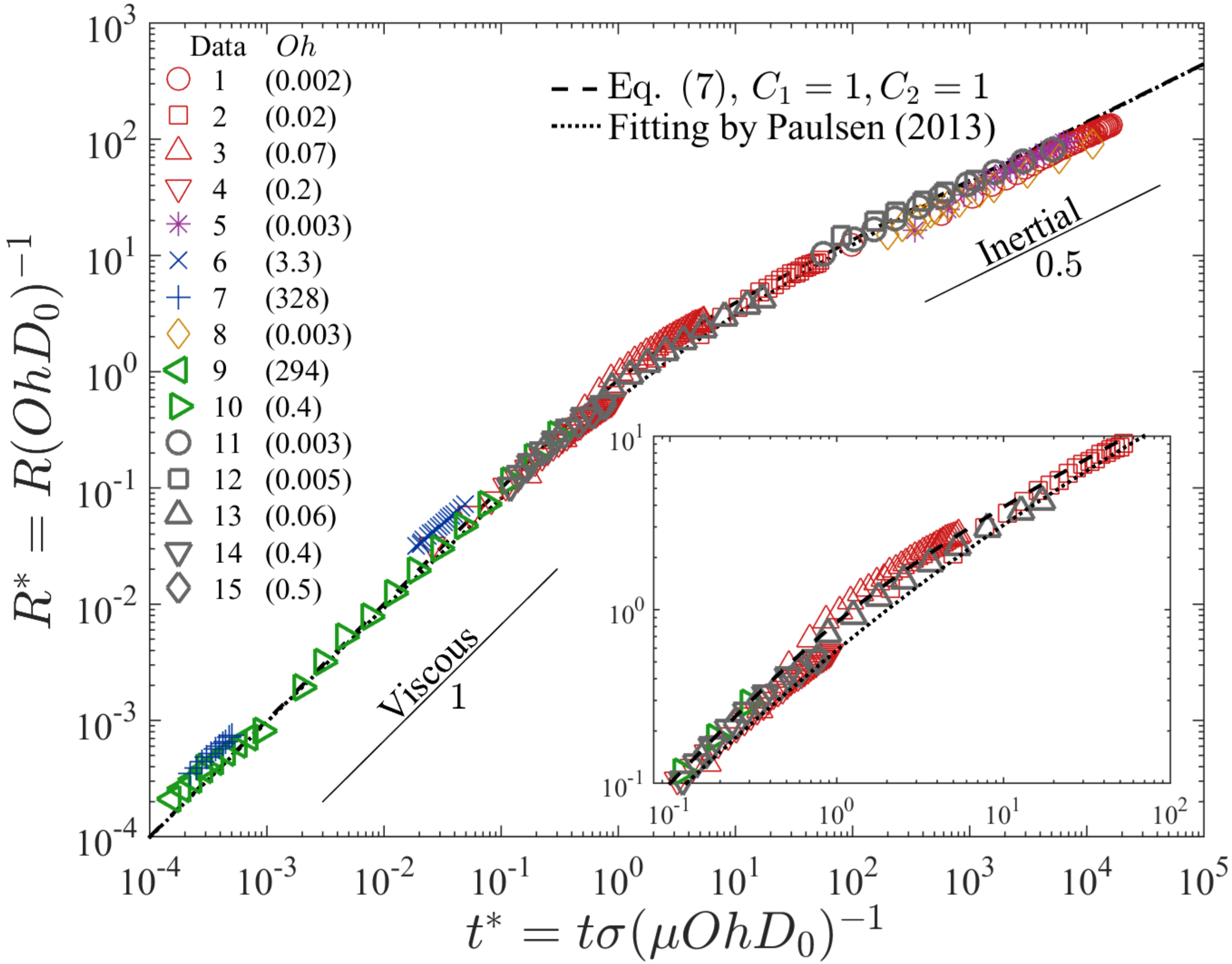}
\caption{Validation of Eq.~(\ref{eq:9}) against experimental data from previous studies \cite{Aarts2005,Thoroddsen2005,Yao2005,Fezzaa2008,Paulsen2013,Paulsen2012}. The solid line is the fitting of all data using the current model of Eq.~(\ref{eq:9}). The dashed line corresponds to this model with asymptotic behaviors in the viscous and inertial regimes identical to the fitting relation of Paulsen \cite{Paulsen2013}, which is represented by the dotted line. }
\label{fig:2}
\end{figure}
%

We now provide the validation of Eq.~(\ref{eq:9}) in Fig.~\ref{fig:2}, by comparing our theory with previous experimental data of various $Oh$. 
The relevant parameters ($D_0$, $\rho$, $\sigma$, and $\mu$) for the 15 sets of experimental data are summarized in the supplemental materials \cite{SM}. It is observed from Fig.~\ref{fig:2} that all data tend to collapse onto a single curve, while the scatteredness of the data reflects the variations of the scaling prefactors reported by different experiments. The collapsed data agree well with the dashed line, which is the current model with $C_1=1$ and $C_2=1$, demonstrating a good universality of this model in predicting the neck evolution, even though the original experiments were conducted under different conditions for the droplets of distinct liquids. 

Fig.~\ref{fig:2} also demonstrates the asymptotic behaviors in the viscous and inertial regimes. Specifically, the $R^{\ast} \sim t^{\ast}$ and $R^{\ast} \sim \sqrt{t^{\ast}}$ scaling relations show up as $R^{\ast} \ll 1$ and $R^{\ast} \gg 1$, respectively, whereas a clear inflection point can be identified around $R^{\ast} = 1$ and $t^{\ast} = 1$, marking the transition from viscous to inertial. The dotted line represents Paulsen's fitting \cite{Paulsen2013}, the functionality of which has been adjusted to $2(R^{\ast})^{-1} = (t^{\ast}/2)^{-1} + (t^{\ast}/2)^{-1/2}$ under the current scales of L and T. Comparing Paulsen's fitting with the current theory ($C_1 = 1$ and $C_2 = 1$), we find that the two curves overlap with each other in the limit $R^{\ast} \ll 1$ and $R^{\ast} \gg 1$. 
The inset plot of Fig.~\ref{fig:2} is a close-up of the crossover regime, showing that Paulsen's fitting in this regime is slightly off compared with our analytical solution. Better fittings by using Eq.~(\ref{eq:9}) and different $C_1$ and $C_2$ can be obtained, but they make overall subtle differences. 

The discussions of Eqs.~(\ref{eq:10}) and~(\ref{eq:11}) together with Fig.~\ref{fig:2} imply that the viscous, viscous-to-inertial crossover, and inertial regimes can be roughly segmented as $R^{\ast} \ll 1$, $R^{\ast} \sim O(1)$, and $R^{\ast} \gg 1$. With $R^{\ast} = R/(Oh D_0)$, the coalescence regime is clearly dictated by two factors, $R/D_0$ and $Oh$. The effect of $R/D_0$ can be understood that for a given fluid the droplet coalescence regime changes from viscous to inertial or has such a trend as the neck expands out naturally. However, $R/D_0$ can only increase up to the order of unity, so it is $Oh$ that eventually decides whether a coalesced droplet could enter the inertial regime. This is evident from Fig.~\ref{fig:2} that data in the inertial regime generally corresponds to smaller $Oh$ and vice versa, given the sampling range of $R/D_0$ does not vary dramatically among different experiments. This contributes to an important understanding of how drop coalescence should be classified in practice. For example, Aarts \textit{et al.} \cite{Aarts2005} used Data 3 (20 mPa s silicon oil) and Data 4 (50 mPa s silicon oil) to demonstrate the inertial scaling, whereas the current study clearly shows that Data 3 mainly covers the crossover regime and Data 4 extends from the viscous regime to the crossover regime. 

\begin{figure}
\includegraphics[scale=0.37]{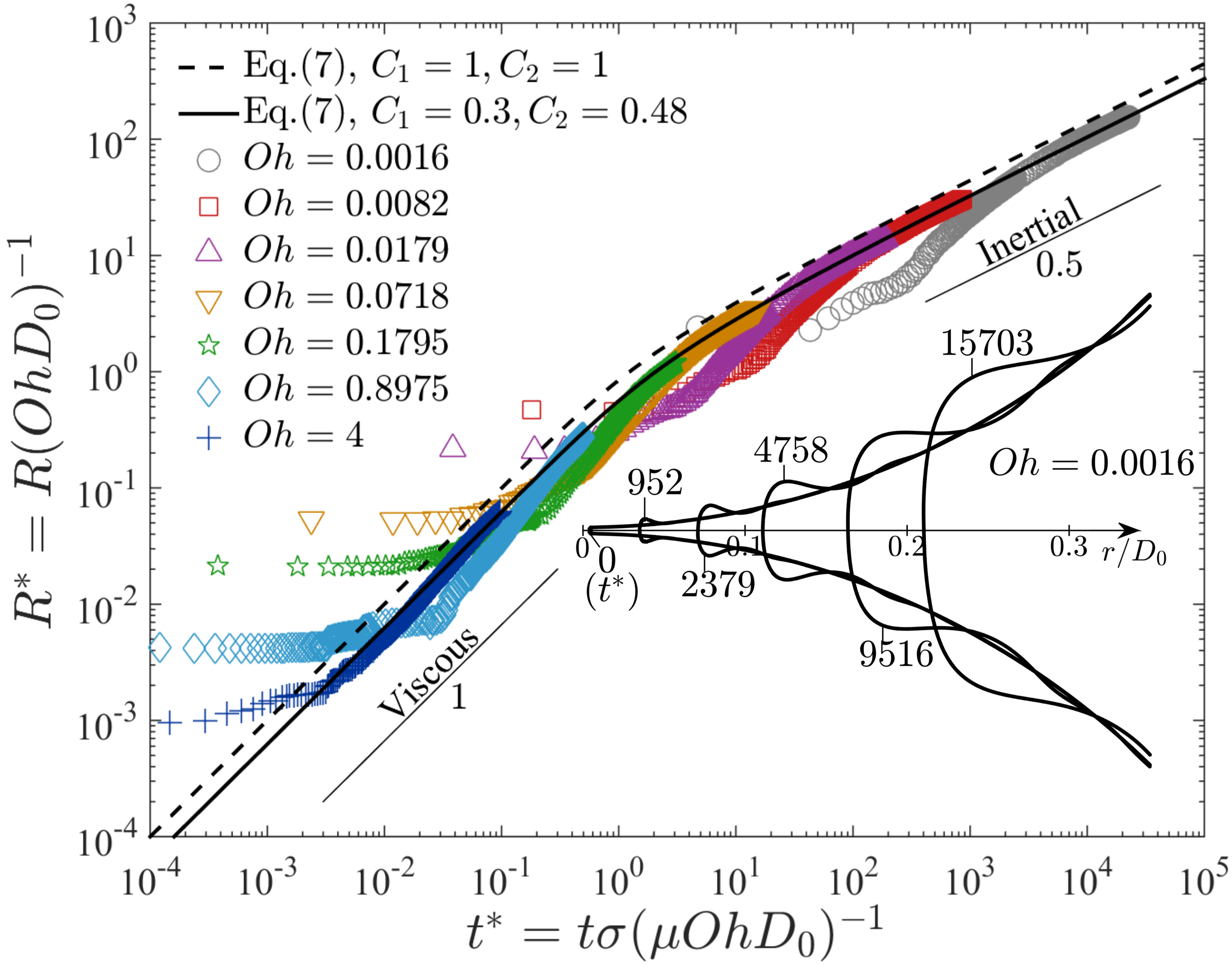}
\caption{Main: validation of Eq. (7) against simulated neck evolution for droplets of different viscosities ($Oh$). Inset: time evolution of the simulated neck interface for a representative case with $Oh = 0.0016$.}
\label{fig:3}
\end{figure}

As an additional support to our theory and the assertion on $Oh$ effects, a volume of fluid (VOF) simulation \cite{Scardovelli1999,Tryggvason2006} was implemented using the open source code, Gerris \cite{Popinet2003,Popinet2009}. This numerical approach has been demonstrated to be suitable for multiphase flow \cite{Chen2014a,Chen2014b,Agbaglah2015,Thoraval2016}, and validated for the droplet coalescence problem in our previous study \cite{Xia2017}. Since the numerical interface is represented by finite layers of grid cells, it inevitably introduces a finite neck radius after the initial contact of droplets. By taking advantage of the adaptive mesh refinement \cite{Popinet2003,Popinet2009}, we have brought the initial neck radius, $R/D_0$, down to the order of $O(10^{-3})$, which would greatly improve the neck evolution in the initial stage. The evolution of the neck interface for a representative simulation case, with $Oh = 0.0016$, is shown in the inset plot of Fig.~\ref{fig:3}. 
Similar simulations were conducted for $Oh=$ 0.0082, 0.0179, 0.0718, 0.1795, 0.8975, and 4. The neck radius evolutions are shown in the main plot of Fig.~\ref{fig:3}. It is seen that each simulation data set originates from a finite neck radius, causing the simulated evolution to deviate from experiment or theory, especially at the very early stages when $R^{\ast}$ hardly grows with time. Nevertheless, the later-stage coalescence behavior is less affected by the simulation onset, as each neck evolution curve gradually approaches and then follows its designated scaling. Similar neck evolution behaviors were also observed from the simulation studies of Sprittles and Shikhmurzaev \cite{Sprittles2012,Sprittles2014}. Based on fitting the late-stage curves of the simulated neck evolution to Eq.~(\ref{eq:9}), we estimated $C_1 = 0.3$ and $C_2 = 0.48$, which is in reasonable agreement with $C_1=1$ and $C_2=1$ obtained based on experiment. The differences in those coefficients are likely caused by the initial condition of the simulation, $R^{\ast}(t^{\ast}=0) \neq 0$, being different from the theory. Last, Fig.~\ref{fig:3} also justifies that both increasing $R/D_0$ and decreasing $Oh$ would cause the coalescence to move towards the more inertial regime, which is consistent with the experimental observation from Fig.~\ref{fig:2}.

To summarize, we have presented a theoretical model for the neck evolution during initial coalescence of binary liquid droplets. We showed that the length and time scales for the crossover between viscous and inertial regimes come naturally in the theory. With the proposed scaling, we derived and validated a unified solution that applies to the viscous, viscous-to-inertial crossover, and inertial regimes of droplet coalescence. This provides a fundamental framework to support the prominent scaling laws as well as the crossover behaviors observed from previous experimental and numerical studies.  

This work was supported by the Hong Kong RGC/GRF (PolyU 152217/14E and PolyU 152651/16E) and partly by the “Open Fund” of State Key Laboratory of Engines (Tianjin University, No. K2018-12).


\begin{thebibliography}{99}

  \bibitem{Yarin2006}
A.L.~Yarin, Annu. Rev. Fluid Mech. {\bf 38}, 159  (2006).

  \bibitem{Gorokhovski2008}
M.~Gorokhovski and M.~Herrmann, Annu. Rev. Fluid Mech. {\bf 40}, 343 (2008).

  \bibitem{Kavehpour2015}
H.P.~Kavehpour, Annu. Rev. Fluid Mech. {\bf 47}, 245 (2015).

  \bibitem{Low1982}
T.B.~Low and R.~List, J. Atmos. Sci. {\bf 39}, 1591 (1982).

  \bibitem{Chen2007}
R.-H.~Chen, Appl. Therm. Eng. {\bf 27}, 604 (2007). 

  \bibitem{Derby2010}
B.~Derby, Annu. Rev. Mater. Res. {\bf 40}, 395 (2010).

  \bibitem{Tang2016}
C.~Tang, J.~Zhao, P.~Zhang, C.K.~Law, and Z.~Huang, J. Fluid Mech. {\bf 795}, 671 (2016).

  \bibitem{Xia2017}
X.~Xia, C.~He, D.~Yu, J.~Zhao, and P.~Zhang, Phys. Rev. Fluids {\bf 2}, 113607 (2017). 

  \bibitem{vandeVorst1994}
G.A.L.~van de Vorst, Technische Universiteit Eindhoven (1994). 

  \bibitem{Dreher1999}
T.M.~Dreher, J.~Glass, A.J.~O'Connor, and G.W.~Stevens, AIChE J. {\bf 45}, 1182 (1999). 

  \bibitem{Squires2005}
T.M.~Squires and S.R.~Quake, Rev. Mod. Phys. {\bf 77}, 977 (2005).

  \bibitem{Frenkel1945}
J.~Frenkel, J. Phys. (Moscow) {\bf 9}, 385 (1945).

  \bibitem{Hopper1993a}
R.W.~Hopper, J. Am. Ceram. Soc. {\bf 76}, 2947 (1993).

  \bibitem{Hopper1990}
R.W.~Hopper, J. Fluid Mech. {\bf 213}, 349 (1990).

  \bibitem{Hopper1993b}
R.W.~Hopper, J. Am. Ceram. Soc. {\bf 76}, 2953 (1993). 

  \bibitem{Eggers1999}
J.~Eggers, J.R.~Lister, and H.A.~Stone, J. Fluid Mech. {\bf 401}, 293 (1999). 

  \bibitem{Duchemin2003}
L. Duchemin, J. Eggers, and C. Josserand, J. Fluid Mech. {\bf 487}, 167 (2003).

  \bibitem{Aarts2005}
D.G.A.L.~Aarts, H.N.W.~Lekkerkerker, H.~Guo, G.H.~Wegdam, and D.~Bonn, Phys. Rev. Lett. {\bf 95}, 164503 (2005).

  \bibitem{Thoroddsen2005}
S.T.~Thoroddsen, K.~Takehara, and T.G.~Etoh, J. Fluid Mech. {\bf 527}, 85 (2005).

  \bibitem{Yao2005}
W.~Yao, H.J.~Maris, P.~Pennington, and G.M.~Seidel, Phys. Rev. E {\bf 71}, 016309 (2005).

  \bibitem{Case2008}
S.C.~Case and S.R.~Nagel, Phys. Rev. Lett. {\bf 100}, 084503 (2008).

  \bibitem{Fezzaa2008}
K.~Fezzaa and Y.~Wang, Phys. Rev. lett. {\bf 100}, 104501 (2008).

  \bibitem{Paulsen2011}
J.D.~Paulsen, J.C.~Burton, and S.R.~Nagel, Phys. Rev. Lett. {\bf 106}, 114501 (2011).

  \bibitem{Wu2004}
M.~Wu, T.~Cubaud, and C.-M.~Ho, Phys. Fluids {\bf 16}, L51 (2004).

  \bibitem{Burton2007}
J.C.~Burton and P.~Taborek, Phys. Rev. Lett. {\bf 98}, 224502 (2007).

  \bibitem{Case2009}
S.C.~Case, Phys. Rev. E {\bf 79}, 026307 (2009).

  \bibitem{Eiswirth2012}
R.T.~Eiswirth, H.J.~Bart, A.A.~Ganguli, and E.Y.~Kenig, Phys. Fluids {\bf 24}, 062108 (2012).

  \bibitem{Pothier2012}
J.C.~Pothier and L.J.~Lewis, Phys. Rev. B {\bf 85}, 115447 (2012).

  \bibitem{Gross2013}
M. Gross, I. Steinbach, D. Raabe, and F. Varnik, Phys. Fluids {\bf 25}, 052101 (2013).

  \bibitem{Sprittles2012}
J.E.~Sprittles and Y.D.~Shikhmurzaev, Phys. Fluids {\bf 24}, 122105 (2012).

  \bibitem{Ristenpart2006}
W.D.~Ristenpart, P.M.~McCalla, R.V.~Roy, and H.A.~Stone, Phys. Rev. Lett. {\bf 97}, 064501 (2006).

  \bibitem{Sanchez2012}
J.F.~Hern\'{a}ndez-S\'{a}nchez, L.A.~Lubbers, A.~Eddi, and J.H.~Snoeijer, Phys. Rev. Lett. {\bf 109}, 184502 (2012).

  \bibitem{Eddi2013}
A.~Eddi, K.G.~Winkels, and J.H.~Snoeijer, Phys. Rev. Lett. {\bf 111}, 144502 (2013).

  \bibitem{Paulsen2013}
J.D.~Paulsen, Phys. Rev. E {\bf 88}, 063010 (2013).

  \bibitem{Tryggvason2011}
G.~Tryggvason, R.~Scardovelli, and S.~Zaleski, \textit{Direct Numerical Simulations of Gas-Liquid Multiphase Flows} (Cambridge University Press, Cambridge, 2011).

  \bibitem{Barenblatt1996}
G.I.~Barenblatt, \textit{Scaling, self-similarity, and intermediate asymptotics} (Cambridge University Press, Cambridge, 1996).

  \bibitem{Paulsen2012}
J.D.~Paulsen, J.C.~Burton, S.R.~Nagel, S.~Appathurai, M.T.~Harris, and O.A.~Basaran, Proc. Natl. Acad. Sci. USA {\bf 109}, 6857 (2012).

  \bibitem{SM}
See Supplemental Materials for a detailed derivation of the velocity integral in \cite{Eggers1999} and relevant parameters of previous experimental data.

  \bibitem{Scardovelli1999}
R.~Scardovelli and S.~Zaleski, Annu. Rev. Fluid Mech. {\bf 31}, 567 (1999).

  \bibitem{Tryggvason2006}
G.~Tryggvason, A.~Esmaeeli, J.~Lu, and S.~Biswas, Fluid Dyn. Res. {\bf 38}, 660 (2006).

  \bibitem{Popinet2003}
S.~Popinet, J. Comput. Phys. {\bf 190}, 572 (2003).

  \bibitem{Popinet2009}
S.~Popinet, J. Comput. Phys. {\bf 228}, 5838 (2009).

  \bibitem{Chen2014a}
X.~Chen and V.~Yang, J. Comput. Phys. {\bf 269}, 22 (2014).

  \bibitem{Chen2014b}
X.~Chen and V.~Yang, Phys. Fluids {\bf 26}, 102104 (2014).

  \bibitem{Agbaglah2015}
G.~Agbaglah, M.-J.~Thoraval, S.T.~Thoroddsen, L.V.~Zhang, K.~Fezzaa, and R.D.~Deegan, J. Fluid Mech. {\bf 764}, R1 (2015).

  \bibitem{Thoraval2016}
M.-J.~Thoraval, Y.~Li, and S.T.~Thoroddsen, Phys. Rev. E {\bf 93}, 033128 (2016).

  \bibitem{Sprittles2014}
J.E.~Sprittles and Y.D.~Shikhmurzaev, J. Fluid Mech. {\bf 751}, 480 (2014).

\end{thebibliography}
\end{document}